\documentclass[a4paper,11pt]{article}
\usepackage{graphicx, rotating,amsmath}
\usepackage{ifpdf}
\ifpdf
\usepackage{hyperref, pdfsync, epstopdf}	
\else
\usepackage[dvips,bookmarks]{hyperref}	
\fi
\hypersetup{colorlinks,bookmarksopen,bookmarksnumbered,citecolor=verdes,
linkcolor=blus,pdfstartview=FitH,urlcolor=rossos}
\def\hhref#1{\href{http://arxiv.org/abs/#1}{#1}} 
\newcommand{\eq}[1]{~{\rm (\ref{eq:#1})}}

\newcommand{\GeV}{\,{\rm GeV}}

\def\circa#1{\,\raise.3ex\hbox{$#1$\kern-.75em\lower1ex\hbox{$\sim$}}\,}

\newcommand{\beq}{\begin{equation}}
\newcommand{\eeq}{\end{equation}}
\newcommand{\beqa}{\begin{eqnarray}}
\newcommand{\eeqa}{\end{eqnarray}}

\font\tenrsfs=rsfs10 at 11pt
\font\sevenrsfs=rsfs7
\font\fiversfs=rsfs5
\newfam\rsfsfam
\textfont\rsfsfam=\tenrsfs
\scriptfont\rsfsfam=\sevenrsfs
\scriptscriptfont\rsfsfam=\fiversfs
\def\mathscr#1{{\fam\rsfsfam\relax#1}}

\def\circa#1{\,\raise.3ex\hbox{$#1$\kern-.75em\lower1ex\hbox{$\sim$}}\,}
\makeatletter
\newcommand{\fig}[1]{~\ref{fig:#1}}
\def\art{\@ifnextchar[{\eart}{\oart}}
\def\eart[#1]#2#3#4#5#6{{\rm #2}, {#3 #4} {\rm (#6) #5} [{\hhref{#1}}]}
\def\hepart[#1]#2{{\rm #2, \hhref{#1}}}
\newcommand{\oart}[5]{{\rm #1}, {\em #2 \rm #3} {\rm (#5) #4}}

\newcounter{alphaequation}[equation]
\def\thealphaequation{\theequation\hbox to
0.6em{\hfil\alph{alphaequation}\hfil}}
\def\eqnsystem#1{
\def\@eqnnum{{\rm (\thealphaequation)}}
\def\@@eqncr{\let\@tempa\relax \ifcase\@eqcnt \def\@tempa{& & &} \or
  \def\@tempa{& &}\or \def\@tempa{&}\fi\@tempa
  \if@eqnsw\@eqnnum\refstepcounter{alphaequation}\fi
\global\@eqnswtrue\global\@eqcnt=0\cr}
\refstepcounter{equation} \let\@currentlabel\theequation \def\@tempb{#1}
\ifx\@tempb\empty\else\label{#1}\fi
\refstepcounter{alphaequation}
\let\@currentlabel\thealphaequation
\global\@eqnswtrue\global\@eqcnt=0 \tabskip\@centering\let\\=\@eqncr
$$\halign to \displaywidth\bgroup \@eqnsel\hskip\@centering
$\displaystyle\tabskip\z@{##}$&\global\@eqcnt\@ne
\hskip2\arraycolsep\hfil${##}$\hfil& \global\@eqcnt\tw@\hskip2\arraycolsep
$\displaystyle\tabskip\z@{##}$\hfil
\tabskip\@centering&\llap{##}\tabskip\z@\cr}
\def\endeqnsystem{\@@eqncr\egroup$$\global\@ignoretrue} \makeatother

\newcommand{\eV}{\,{\rm eV}}

\usepackage{multicol}
\usepackage{color}
\definecolor{rosso}{cmyk}{0,1,1,0.4}
\definecolor{rossos}{cmyk}{0,1,1,0.55}
\definecolor{rossoc}{cmyk}{0,1,1,0.2}
\definecolor{blu}{cmyk}{1,1,0,0.3}
\definecolor{blus}{cmyk}{1,1,0,0.6}
\definecolor{bluc}{cmyk}{1,1,0,0.1}
\definecolor{verde}{cmyk}{0.92,0,0.59,0.25}
\definecolor{verdec}{cmyk}{0.92,0,0.59,0.15}
\definecolor{verdes}{cmyk}{0.92,0,0.59,0.4}

\begin{document}
\begin{center}

\bigskip\bigskip

\color{black}
{\Huge\bf\color{rossos} New bounds on millicharged particles from cosmology}
\medskip
\bigskip\color{black}\vspace{0.5cm}

{{\large\bf A.\ Melchiorri}$^a$,
{\large\bf A.D.\ Polosa}$^b$,
{\large\bf A.\ Strumia}$^c$
}
\\[7mm]
{\it $^a$  Dipartimento di Fisica, Universit\`a di Roma I
Piazzale A.\ Moro 2, I-00185 Roma Italy}\\
{\it $^b$  INFN Roma I,
Piazzale A.\ Moro 2, I-00185 Roma Italy}\\
{\it $^c$ Dipartimento di Fisica dell'Universit{\`a} di Pisa and INFN, Pisa, Italia}\\
\end{center}

\bigskip

\centerline{\large\bf\color{blus} Abstract}

\begin{quote}\large\color{blus}
Particles with millicharge $q$ and sub-eV mass can be produced
in photon-photon collisions, 
distorting the energy spectrum of the Cosmic Microwave Background. 
We derive the conservative bound $q\circa{<} 10^{-7}e$
(as well as model-dependent bounds two orders of magnitude stronger),
incompatible with proposed interpretations of the PVLAS anomaly based
on millicharged production or on millicharged-mediated axion-like couplings.


\color{black}
\end{quote}

\section{Introduction}

New particles $q$ with mass $m_q$ and charge $q\ll e$ (thereby called `millicharged')
can naturally arise in field-theory models, e.g.\
via small kinetic mixing of the photon with a new vector.
Distinctive aspects and constraints on millicharged particles are reviewed 
in~\cite{Davidson}.

Renewed interest in millicharged particles was prompted by an experimental result
claimed by the PVLAS collaboration~\cite{pvlas}: 
the polarization of a linearly polarized
laser beam  (photons with $E_\gamma\sim 1\eV$) rotates (`dichroism') and develops
a tiny elliptical component (`birefringence') after multiple passages 
trough a vacuum Fabry-Perot cavity containing a (rotating) $5.5$~Tesla magnetic field orthogonal to the beam direction.
If confirmed, PVLAS calls for new physics, possibly pointing to the existence of new light particles that, interacting with photons, drain part of the beam energy in a polarization-dependent way.

Thanks to the external field, photons in the beam can non-perturbatively
convert into couples of new
light millicharged particles: one can fit PVLAS for 
charge $q\sim 3 \cdot 10^{-6}e$ and mass $m_q\sim 0.1\eV$~\cite{Ringwald,Foot}.
Alternatively, the PVLAS birefringence could be due to mixing, induced by the magnetic field,  
of laser photons with a new axion-like scalar $a$, coupled to photons by an effective interaction term 
$(a/4M)\, F_{\mu\nu} F^{\mu\nu}$
(a pseudo-scalar axion coupled as $(a/4M)\, F_{\mu\nu}\tilde{F}^{\mu\nu}$
produces birefringence with sign opposite to the one observed by PVLAS)~\cite{mpz},
and the PVLAS dichroism could be due to conversion of some photons in axions escaping from the apparatus.
The needed value of $M \sim 2\cdot 10^5 \GeV$ is excluded by $a$ production in stars~\cite{cast},
unless the effective coupling is mediated by a loop of light enough particles,
$1/M \sim \alpha q^2/v$, that therefore need to have 
a small millicharge $q \approx 10^{-6} e\cdot \sqrt{v/\eV}$, 
where $v$ is a loop function, related and comparable to $m_q$~\cite{Redondo}.\footnote{Other interpretations not involving millicharged particles have been put forward. 
Ref.~\cite{antoniadis} suggested that the axion-like scalar might be replaced with
a massive vector $B_\mu$ with a Chern-Simons-like coupling to photons $A_\mu$.
However we do not see how the $A_\mu \leftrightarrow B_\mu$ oscillations proposed in~\cite{antoniadis}
can produce
the dichroism claimed by PVLAS.
It could be produced by emission of 
the longitudinal component of the extra vector, with couplings enhanced
by the polarization vector
$\epsilon_\mu \sim {q_\mu}/{m}$, but (just like in the axion case) this
possibility is not compatible with bounds from star cooling. 
Extra observables related to axion-like physics are discussed in~\cite{mimmo}.}

Light millicharged particles therefore are a key ingredient of 
proposed new physics that can fit the PVLAS anomaly.
Still, one needs to  circumvent the too strong constraints on millicharged  particles in~\cite{Davidson,} 
by designing models that exploit the fact that PVLAS involves ${\cal O}(\eV)$ energies,
while the constraints in~\cite{Davidson} come from physics at
${\cal O}({\rm keV})$ energies (the temperature in the core of stars) or higher.
An energy-dependent form factor 
that suppresses millicharged couplings at high energies can be obtained in the following way.
(Although we use a basis that makes its presentation simpler, we are just outlining
the model of~\cite{Redondo}).
The gauge group is ${\rm U}(1)_{\rm em}\otimes{\rm U}(1) \otimes{\rm U}(1)'$
where the first U(1)$_{\rm em}$ is the usual electromagnetism, and new light particles
(either scalars or fermions)  are charged
under the last U(1)$'$, with charge $q'$, assumed to be comparable to $e$~\cite{Redondo}.
The propagator matrix of the three vectors is assumed to be~\cite{Redondo}
\beq\label{eq:matrix}
\begin{pmatrix}
& q^2 & \epsilon q^2 & 0 \\
&\epsilon q^2 & q^2 + m^2 & m^2 \cr
&0 & m^2 & q^2 + m^2
\end{pmatrix}\eeq
i.e.\ canonical kinetic terms, plus a small kinetic mixing between the first two U(1),
plus a vev that breaks the last two U(1) to their diagonal sub-group, 
giving a mass $\sim m$ to their symmetric combination.
At $q^2 \ll m^2$, the new light particles get a millicharge $q \sim \epsilon q'$ under the photon.
At $q^2 \gg m^2$ one can neglect $m^2$ and this millicharge disappears:
choosing $m\circa{<} \eV$ allows to avoid the bounds from higher-energy physics discussed in~\cite{Davidson}.\footnote{The Higgs $h^\prime$ that breaks ${\rm U}(1) \otimes{\rm U}(1)'$ 
to its diagonal subgroup
providing the vector mass $m$
gets a millicharge $q\sim \epsilon q'$ with no form factor that suppresses it at high energy:
therefore the model is excluded if $h^\prime$ is light enough to be produced in stars or during BBN.
Since $m_{h^\prime} \sim \lambda m/g^\prime$, a heavy enough $h^\prime$ needs a quartic scalar coupling $\lambda |h'|^4$ with $\lambda \gg 1 $.
This may appear crazy, and indeed such a non-perturbative Higgs coupling would be excluded
if $h'$ would break a non-abelian gauge group, because non-abelian vectors feel it
(for example the SM Higgs must be lighter than $4\pi M_Z/g\sim {\rm TeV}$).
However the model only involves abelian vectors, and
in the abelian case a big $\lambda$ remains confined to the Higgs sector, 
so that decoupling the physical Higgs is not impossible.
Indeed this is well known~\cite{Stuckelberg}: an abelian theory remains sensible if
gauge invariance is broken by
adding a vector mass term (i.e.\ no Higgs or infinitely heavy Higgs).
In conclusion, the models of~\cite{Redondo,Ringwald} an unusual but 
acceptable ingredient.}

Searches for this kind of millicharged particles have been performed using  reactor neutrino experiments,
obtaining the bound $q\circa{<} 10^{-5}e$~\cite{Rubbia}.
New reactor experiments have been discussed as a way to test interpretations of the PVLAS anomaly that involve millicharged particles~\cite{Rubbia}.

\bigskip

We show, trying to be as model-independent as possible, that 
existing models are not compatible with known CMB physics after recombination,
which probes the same sub-eV energies relevant for PVLAS,
and we obtain new bounds on millicharged particles.


\section{Model-independent cosmological constraints}

Millicharged particles affect CMB cosmology in many ways.
We list the relevant processes, discuss their rate, their effect,
and their model-independence.
We restrict our attention to light millicharged particles, $m_q\circa{<}\eV$,
and on cosmology after decoupling, $T\circa{<}\eV$:
according to the SM, photons behave as free particles with negligible thermal mass:
their dispersion relation is $q^2 =0$.
We conservatively assume that the universe initially contains only ordinary SM particles. 
\begin{itemize}
\item[a)] $\gamma\gamma \to q \bar{q}$ is the main process for production of millicharged particles.
Being induced by a small adimensional gauge coupling, this process is maximally effective at low
temperatures, $T_* \sim \max(m_q,T_0)$
(where $T_0 =2.7\,{\rm K} = 0.23\,{\rm meV}$ is the present temperature), 
so that its rate is dominated by
model-independent gauge interactions, controlled only by $m_q$ and $q$.\end{itemize}
We can neglect the process $\gamma e\to e q\bar{q}$,
that contains Bethe-Heitler-like contributions $\sigma \sim e^2 q^4/(4\pi)^3T_*^2$
(suppressed with respect to $\gamma\gamma \to q\bar{q}$ by a
factor $\sim e^2/(4\pi)^2$) and
Compton-like contributions $\sigma \sim e^4 q^2/(4\pi)^3 m_e^2$
(suppressed by the electron mass).
Both contributions give rates additionally suppressed by the small number
density of free electrons, $n_e/n_\gamma\ll 1$.

We can estimate the number density $n_q$ of millicharged
particles plus their anti-particles produced by process a) as
\beq Y\equiv \frac{n_q}{n_\gamma}
 \sim \min\left(1,\sigma(\gamma\gamma \to q \bar{q}) \frac{ n_\gamma(T_*)}{H(T_*)}\right)\qquad \hbox{where}\qquad
 \sigma(\gamma\gamma \to q \bar{q}) \sim \frac{q^4}{4\pi T_*^2}\eeq
 and  $n_\gamma(T) = 2\zeta(3) T^3/\pi^2$ is the photon number density,
 used to normalize $n_q$ such that  their relative density $Y= n_q/n_\gamma$ is not affected by the expansion, with Hubble rate $H(T) = \dot R/R$.

\begin{figure}[t]
\begin{center}
\includegraphics[width=0.9\textwidth]{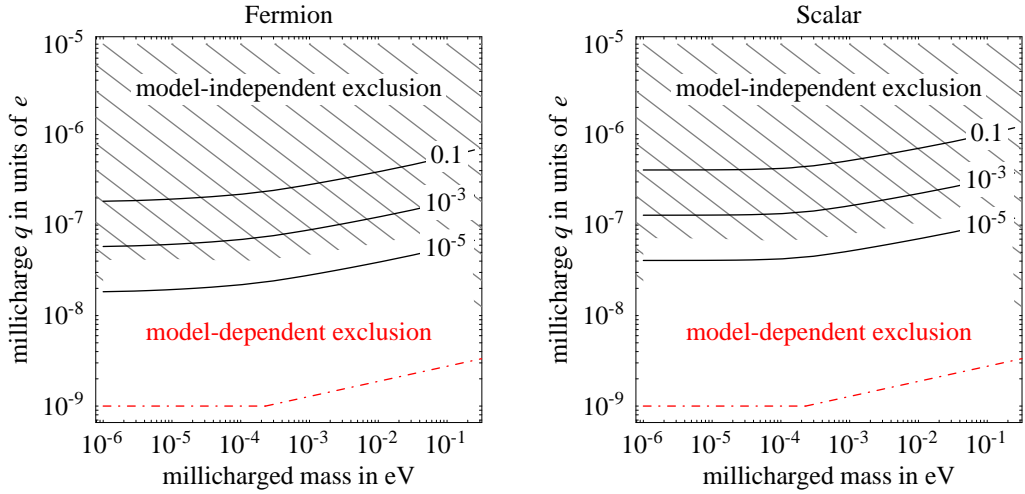}
\caption{\label{fig:nq}\em
Isocurves of the cosmological abundance $n_q/n_\gamma$ 
of fermionic (left plot)
or scalar (right plot) millicharged particles as function of their mass $m_q$ and charge $q$.
The shaded region is excluded by the CMB energy spectrum.
The dot-dashed 
curve is the (robust, but not fully model-independent) constraint estimated in  eq.\eq{Yd}, for
$q'\sim e$.
For comparison, the PVLAS anomaly can be interpreted as production of millicharges with
$q\sim \hbox{\rm few}\times10^{-6}e$
and maybe $m_q\sim 0.1\eV$~\cite{Ringwald}.
}
\end{center}
\end{figure}

\begin{figure}[t]
\begin{center}
\includegraphics[width=0.8\textwidth]{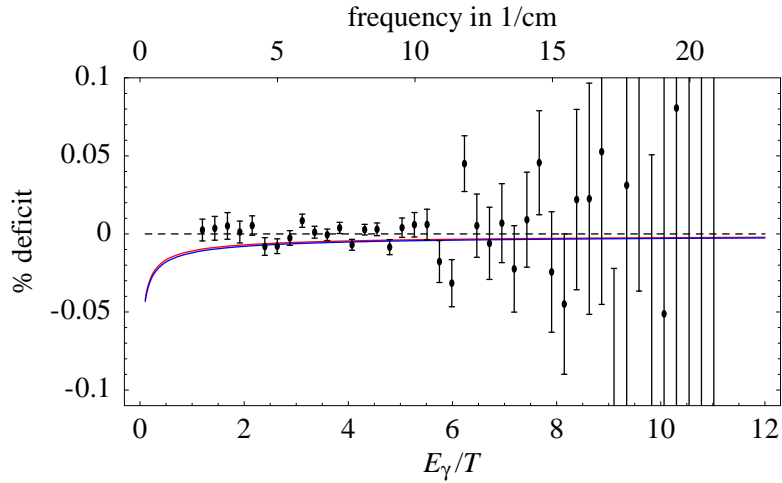}
\caption{\label{fig:spettro}\em FIRAS data compared to
the energy-dependent depletion  of the CMB spectrum due to
$\gamma\gamma\to q\bar{q}$.
We plot $1-r$, with $r$ given in eq.\eq{r}, 
computed for fermion (scalar) millicharges with
$m_q = 0.1\eV$ and
$q=10^{-7}e$ ($q=1.7~10^{-7}e$), chosen such that the two cases give roughly equal effects,
excluded at about 3 standard deviations.}
\end{center}
\end{figure}

Fig.\fig{nq} shows isocurves of $Y$, more precisely computed solving the Boltzmann
equations described in the Appendix.
If $Y\ll 1$ the process
 $\gamma\gamma\to q\bar{q}$ leads to an energy-dependent depletion of the CMB,
 whose energy spectrum has been measured with $\sim 10^{-5}$ accuracy by FIRAS~\cite{FIRAS}. 
 As  illustrated in fig.\fig{spettro}, by fitting FIRAS data we find
\beq\label{eq:Ybound}
Y = n_q/n_\gamma \circa{<}
6\cdot 10^{-5} \qquad\hbox{at $3\sigma$ confidence level ($\Delta \chi^2 = 9$)}.\eeq
for $m_q\sim 0.1\eV$, and a slightly weaker bound at smaller $m_q$.
Fig.\fig{nq} shows the precise constraint in the $(q,m_q)$ plane.
We  ignored effects possibly related to the formation of bound states at recombination, 
as they would appear at the frequencies of the H and $^4$He recombination:
a region of the CMB spectrum not tested by FIRAS and
dominated by the galactic signal of thermal dust.

Before concluding that we can safely exclude the interpretation of the PVLAS anomaly based on~\cite{Ringwald}
(the interpretation of~\cite{Redondo} employs values of ($q,m_q$) that can be  
not far from the allowed region) we need to assess if this bound is model-independent,
or if one can add some other new physics and use it to restore a thermal CMB spectrum.

We do not see how additional interactions, slower than the expansion rate, could 
provide such a restoration 
since the CMB spectrum has been measured in a sizable range of energies.
New interactions,   enough faster  than the interaction rate, would re-thermalize photons, but at the expense of thermalizing also some new particle (this happens e.g.\ in the model of~\cite{Foot}):
data on  cosmological anisotropies disfavors the resulting significant depletion
of the photon energy density with respect to other components,
and the addition of new quasi-relativistic interacting particles, 
that behave as a non freely streaming fluid.


This problem is avoided if the new particle, added to keep photons in thermal equilibrium, has a mass $m\gg T_0$, such that
it decouples at $T\circa{<} m$.
In the simplest scenario the millicharged particles themselves could do the job.

Indeed the process a) is much faster than the expansion rate if $q \circa{>}10^{-5}e$; however such a large millicharge cannot fit PVLAS (that actually excludes a too large millicharge) and leads to a thermalized rate also at higher temperatures, thereby 
significantly distorting CMB anisotropies imprinted 
around last scatterings, at $T\approx 0.25\eV$.

Without performing the necessary dedicated analysis, it seems unlikely that
such a big  modification of cosmology could be compatible with observations:
in the tight coupling limit the photon/baryon/millicharged fluid would
have a sound speed different than the usual photon/baryon fluid
(whose acoustic oscillations have been observed as peaks in the 
CMB anisotropy angular spectrum).

Similar problems arise if one instead
assumes that millicharged particles are thermalized at the same
temperature as photons: extra non-freely streaming particles,
photons that remain coupled after recombination,
a non standard amount of relativistic energy density at recombination.
Since interaction rates mediated by adimensional couplings 
become slower than the expansion rate at large temperatures,
this alternative scenario is unnatural and we see
no preferred way that realizes it in a clean way.
One can imagine many different possibilities;
for example the effective number of
neutrinos (commonly used to parameterize the relativistic energy density
at recombination, probed by CMB data)
typically deviates from $3$ by some ${\cal O}({\rm few})$ factor
and can even be smaller than 3,
in scenarios where millicharged particles reheat photons
below the decoupling temperature of neutrinos.


\medskip

Once that enough millicharged particles have been produced, photons start having significant interactions with them.  
Let us now turn to examine the potentially most remarkable of such scattering processes.

\begin{itemize}

\item[b)] $\gamma q \to \gamma q$ leads to a photon attenuation length.
Its cross section is $\sigma =q^4/6\pi m_q^2$ in the non-relativistic Thompson  limit.
This process does not lead to new significant constraints and
can be suppressed in a model-dependent way,
as discussed at point d).




\end{itemize}
Furthermore, there can be model-dependent scatterings involving new-physics particles.
The presence of at least one new extra new light vector $\gamma'$
seems almost model independent;  e.g.\ in the model outlined in the introduction
the millicharged particle has a sizable  gauge coupling $q' \sim e$ to two
extra vectors $\gamma'$ with masses $m'=0$ and $m'\sim m$.
Therefore, unless all $\gamma'$ are too heavy for being relevant, they have sizable effects.

\begin{itemize}
\item[c)]  $\gamma q \to \gamma' q$, with cross section
$\sigma=q^2 q^{\prime 2}/6\pi m_q^2$ in
the limit of non-relativistic millicharge $q$ and of ultra-relativistic $\gamma'$.
If $q'\sim e$ one gets a constraint  orders of magnitude stronger than previous bounds
by demanding that $\gamma q \to \gamma' q$ does not distort the CMB energy spectrum.
(This process can easily be in thermal equilibrium, again restoring a thermal CMB spectrum
but also generating the other problems previously discussed).
Coherent forward scattering, i.e.\ $\gamma \leftrightarrow \gamma'$ oscillations, are
suppressed by a mixing angle $\theta\sim q/q'$
which is small in the most plausible part of the parameter space, $q'\sim e$.



%


\item[d)] Scatterings b) and c) involve an initial state millicharge $q$: 
its density can be suppressed by 
$q\bar{q}\to \gamma'\gamma'$. 
A sizable suppression of $n_q$ is obtained only if  $m_{\gamma'} \circa{<} T \ll m_q$ and if
this process is in thermal equilibrium
(this is indeed the most natural situation, 
since $\sigma(q\bar{q}\to \gamma'\gamma' )\sim q^{\prime 4}/m_q^2$ is not suppressed by any millicharged coupling),
while $n_q$ is only reduced by an ${\cal O}(1)$ factor at $T\sim m_q$.
\end{itemize}
The above considerations suggest another robust (but not fully model-independent) constraint,
coming from the distortion of the CMB energy spectrum due to
$\gamma q\to \gamma' q$ at temperatures $T\sim T_*$.
The resulting total deficit of photons is estimated to be 
\beq   Y_\gamma \sim\min\left(1, \sigma(\gamma q \to \gamma' q) \frac{ Y n_\gamma(T_*)}{H(T_*)}\right)\qquad
\hbox{where}\qquad \sigma(\gamma q \to \gamma' q)\sim \frac{q^2 q^{\prime 2}}{4\pi T_*^2}\eeq
Since this effect is energy-dependent, FIRAS data imply
$Y_\gamma \circa{<} 10^{-4}$ which translates into 
\beq \label{eq:Yd}q'\circa{<} 10^{-9}e ~ \bigg(\frac{q'}{e}\bigg)^{1/3} \max \left(1, \frac{m_q}{T_0} \right) ^{1/6}  .  \eeq
This strong bound depends weakly on the model-dependent parameter $q'$, which is
comparable to $e$ if one wants to get millicharged particles from a small $\gamma/\gamma'$ kinetic
mixing, rather than putting by hand a small gauge coupling.
This bound is grossly incompatible with the values of $q,q'$ used in the model of~\cite{Redondo}
to mediate at one loop the effective operator $aF_{\mu\nu}F^{\mu\nu}/4M$.
Indeed this operator  generates itself a $Y_\gamma \sim 10^{-(7 \hbox{--} 8)}$,
a few orders of magnitude below the sensitivity of FIRAS, 
for the values of  $M\sim \hbox{few} \cdot 10^5\GeV$ and of the axion-like mass $m_a \sim$ few meV
suggested by PVLAS.


\section{Conclusions}
Motivated by the PVLAS anomaly~\cite{pvlas}, the authors of~\cite{Redondo,Ringwald} proposed 
new models containing millicharged particles whose charge only appears at low energy
(or more precisely at small momentum transfer),
avoiding the  bounds  on millicharged particles from higher-energy probes, such as star cooling and BBN~\cite{Davidson}.
Millicharged particles are used to directly fit PVLAS in~\cite{Ringwald}, and to mediate an axion-like coupling to a
new light scalar in~\cite{Redondo}.

Cosmology at low energies (namely at $T\circa{<}\eV$ after recombination)
provides a direct and sensitive probe to this sort of new physics.
Within standard cosmology photons behave as free particles with negligible thermal mass.
New physics processes like $\gamma \gamma \to q \bar{q}$ 
distort the CMB energy spectrum,
potentially conflicting with FIRAS that measured a black-body energy spectrum
up to $\sim10^{-5}$ accuracy~\cite{FIRAS}.
The resulting constraints on millicharged particles, summarized in fig.\fig{nq},
 are therefore based on simple and fully safe cosmology,
 and exclude the interpretations of the PVLAS anomaly proposed in~\cite{Redondo,Ringwald}.
We cannot propose modifications of the models that
avoid the conflict with FIRAS data.



\paragraph{Acknowledgements} 
We thank R. Barbieri, A. Macchi, L. Maiani,  E. Gabrielli, U. Gastaldi, M. Roncadelli  
and especially  S. Davidson for many useful discussions and suggestions.

\small\appendix
\section*{Appendix}

The evolution of $Y = n_q/n_\gamma $ is described by the Boltzmann equation
\beq sHz \frac{dY}{dz} =
  -2\bigg(\frac{Y^2}{Y^{2}_{\rm eq}}-1\bigg)\gamma_A  \,, \eeq
where $z=m_q/T$, $H$ is the expansion rate.
The thermally averaged interaction rate for $\gamma\gamma\leftrightarrow q\bar{q}$ is given by
\beq \label{eq:gammaA}
\gamma_A =  \frac{T}{64 \pi^4} \int_{s_{\rm min}}^{\infty} ds~ s^{1/2}
 {\rm K}_1\bigg(\frac{\sqrt{s}}{T}\bigg)
  \hat{\sigma}(s)\eeq
where the reduced cross section,
defined as $d\hat{\sigma}/dt = \sum |A|^2/8\pi s$ where the sum runs over all initial- and final-state indices, is
\beq \hat\sigma(s) = \frac{q^4}{\pi   x^2}
 \left[-\beta x(4+x)+(x^2+4x-8) \ln\frac{1+\beta}{1-\beta }\right]
 \eeq
 if millicharged particles are a Dirac fermion
 and
 \beq \hat\sigma(s) = \frac{q^4}{2\pi   x^2}
 \left[\beta x(4+x)+4(2-x) \ln\frac{1+\beta}{1-\beta }\right]\eeq
 if millicharged particles have spin 0.
 Here $x\equiv s/M^2$ and $\beta=\sqrt{1-4/x}$ is the $q,\bar{q}$ velocity with respect to their center-of-mass frame. 

 Assuming $Y\ll1$
the Bolztmann equation for the distortion in the photon energy spectrum, 
$r(\epsilon) = f_\gamma(\epsilon)/f_{\rm BE}(\epsilon)$, is
 \beq \label{eq:r}
 Hz \frac{dr(\epsilon)}{dz} =  \frac{1}{32 \pi^2 \epsilon}\int{dc\, d\epsilon^\prime ~ \epsilon^\prime} f_{BE}(\epsilon^\prime)\cdot \hat\sigma(s = 2\epsilon \epsilon^\prime(1-c)T^2)\eeq
 where $\epsilon = E/T$ is the comoving photon energy, $f_{\rm BE}(\epsilon)=1/(e^\epsilon-1)$ is the Bose-Einstein distribution function and $c$ is the cosine of the scattering angle between the two photons.
($Y$ is computed using eq.\eq{gammaA}, 
where, as usual, all thermal distributions are approximated
with the Maxwell distribution: 
this approximation makes almost a factor of 2 difference in $Y$,
that was compensated in eq.\eq{Ybound} by properly 
rescaling the right-handed side).


\end{document}